\documentclass[12pt]{iopart}
\usepackage{graphicx,float,alltt,psfig,curvesls,eepic}
\newcommand{\beq}{\begin{equation}}
\newcommand{\eeq}{\end{equation}}
\newcommand{\bea}{\begin{eqnarray}}
\newcommand{\eea}{\end{eqnarray}}
\newcommand{\bctr}{\begin{center}}
\newcommand{\ectr}{\end{center}}
\newcommand{\llsim}{\mbox{$\:\stackrel{<}{_{\sim}}\:$} }

\newcommand{\omol}{(\Omega_{\rm M},\Omega_{\Lambda})}

\newcommand{\om}{\Omega_{\rm M}}
\newcommand{\oll}{\Omega_{\Lambda}}
\begin{document}

\title{Black hole versus cosmological horizon entropy}
\author{Tamara M. Davis, P. C. W. Davies \& Charles H. Lineweaver}
\begin{abstract}
The generalized second law of thermodynamics states that entropy always increases when all event horizons are attributed with an entropy proportional to their area.  We test the generalized second law by investigating the change in entropy when dust, radiation and black holes cross a cosmological event horizon.  We generalize for flat, open and closed Friedmann-Robertson-Walker universes by using numerical calculations to determine the cosmological horizon evolution.  In most cases the loss of entropy from within the cosmological horizon is more than balanced by an increase in cosmological event horizon entropy, maintaining the validity of the generalized second law of thermodynamics.  However, an intriguing set of open universe models show an apparent entropy decrease when black holes disappear over the cosmological event horizon.  We anticipate that this apparent violation of the generalized second law will disappear when solutions are available for black holes embedded in arbitrary backgrounds.
\end{abstract}
\maketitle

\renewcommand\baselinestretch{1.0}
\normalsize
\section{Introduction}

A significant advance in physical theory was made by Bekenstein with the suggestion (Bekenstein 1970) that the area of the event horizon of a black hole is a measure of its entropy. This hinted at a deep link between information, gravitation and quantum mechanics that remains tantalizingly unresolved today. Bekenstein's claim was bolstered by Hawking's application of quantum field theory to black holes (Hawking 1975), from which he deduced that these objects emit thermal radiation with a characteristic temperature,
\beq T_{\rm b} = \frac{1}{8\pi m_{\rm b}}, \eeq
for a Schwarzschild hole, where $m_{\rm b}$ is the mass of the black hole, and we use units $G = \hbar = c = k = 1$.  Hawking's calculation enabled the entropy of a black hole $S_{\rm b}$ to be determined precisely as,
\bea S_{\rm b} &=& 16\pi m_{\rm b}^2, \\
            &=& \frac{A_{\rm b}}{4}, \label{eq:Sbh}\eea
where $A_{\rm b}$ is the event horizon area.  Eq.~\ref{eq:Sbh} also applies to spinning and charged black holes. It was then possible to formulate a generalized second law of thermodynamics (GSL),
\beq \dot{S}_{\rm env} + \dot{S}_{\rm b} \ge 0, \eeq
where $S_{\rm env}$ is the entropy of the environment exterior to the black hole and an overdot represents differentiation with respect to proper time, $t$. Thus when a black hole evaporates by Hawking radiation its horizon area shrinks, its entropy decreases, but the environment gains at least as much entropy from the emitted heat radiation (Hawking, 1975). 
Conversely, if a black hole is immersed in heat radiation at a higher temperature, radiation will flow into the black hole and be lost. The corresponding entropy reduction in the environment is offset by the fact that the black hole gains mass and increases in area and entropy.

Gibbons \& Hawking (1977) conjectured that event horizon area, including cosmological event horizons, might quite generally have associated entropy.  A prominent example is de Sitter space, a stationary spacetime which possesses a cosmological event horizon at a fixed distance $(3/\Lambda)^{1/2}$ from the observer, where $\Lambda$ is the cosmological constant. It was known (see e.g. Birrell \& Davies 1981) that a particle detector at rest in de Sitter space responds to a de Sitter-invariant quantum vacuum state as if it were a bath of thermal radiation with temperature,
\beq T_{\rm deS} = \frac{1}{2\pi\;\Lambda^{1/2}}. \eeq
It thus seemed plausible that the GSL could be extended to de Sitter space. Subsequent work by Davies (1984), and Davies, Ford and Page (1986) supported this conclusion. There were, however, some problems. Although the de Sitter horizon has thermal properties, the stress-energy-momentum tensor of the de Sitter vacuum state does not correspond to that of a bath of thermal radiation (unlike for the black hole case). Instead, it merely renormalizes the cosmological constant. Secondly, there is no asymptotically flat external spacetime region for de Sitter space, which precludes assigning a mass parameter to the de Sitter horizon. This makes it hard to interpret trading in energy and entropy, as is conventional in thermodynamic considerations, between de Sitter space and an environment. A final problem is that in the black hole case Bekenstein attributed the entropy of the hole to its total hidden information content, which is readily evaluated. For a cosmological horizon, which may conceal a spatially infinite domain lying beyond, the total hidden entropy would seem to be ill-defined. 
Some of the most recent work addressing these issues can be found in Padmanabhan (2002).  

The foregoing concerns are amplified in the case of more general cosmological horizons that are non-stationary and do not even have an associated well-defined temperature. Consider the general class of Friedmann-Robertson-Walker (FRW) models with scalefactor $R(t)$,  
\beq ds^2 = -dt^2+R^2(t)\left[\chi^2+S_k^2(\chi)(d\theta^2+\sin^2\theta d\psi^2)\right],\eeq 
where $S_k(\chi) = \sin\chi, \chi, \sinh\chi$ for closed, flat and open models respectively.   
One may define a conformal vacuum state adapted to the conformally flat geometry of these spaces, and consider the response of a quantum particle detector (Birrell \& Davies 1981, Section 3.3) to such a state. The response will generally be non-zero, but the perceived spectrum will not be thermal. This raises the question: just how far can one extend the GSL to event horizons? Could it apply even to non-stationary cosmological models in spite of the absence of a clear thermal association? And if the GSL cannot be thus extended, what are the criteria that determine the limits of its application? 

We consider these questions to be of significance to attempts to link information, gravitation and thermodynamics, and in recent discussions about the total information content of the universe (Lloyd 2002). They may also assist in attempts to formulate a concept of gravitational entropy
, and to clarify the status of the holographic principle (Susskind, 1995; Bousso 2002).  

In this paper we explore the range of validity of the GSL.  We assume cosmological event horizons do have entropy proportional to their area, as Gibbons and Hawking (1977) proposed. The total entropy of a universe is then given by the entropy of the cosmological event horizon plus the entropy of the matter and radiation it encloses.  In Sect.~\ref{sect:dustFRW} and Sect.~\ref{sect:radFRW} we assess the loss of entropy as matter and radiation disappear over the cosmological event horizon and show that the loss of entropy is more than balanced by the increase in the horizon area.  We then consider in Sect.~\ref{sect:bhFRW} the case of a FRW universe filled with a uniform non-relativistic gas of small black holes.  This enables a direct entropic comparison to be made between black hole and cosmological event horizon area.  As the black holes stream across the cosmological horizon, black hole horizon area is lost, but the cosmological horizon area increases.  We may thus assess the relative entropic `worth' of competing horizon areas.

\section{Dust filled universe}\label{sect:dustFRW}
The simplest case to consider is the classic homogeneous, isotropic FRW universe filled with pressureless dust.  The dust in this model is assumed to be comoving.  The dust is therefore in the most ordered state possible and has zero entropy which allows us to restrict our thermodynamic considerations to the cosmological event horizon alone.  
\begin{figure}
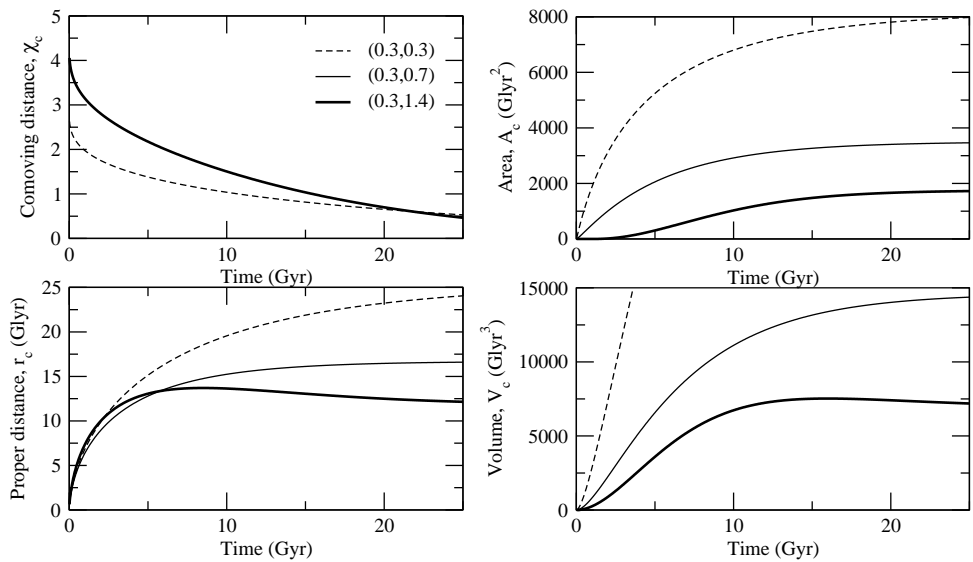
\bctr
\includegraphics[width=61mm]{tXehrehAehVeh5-half1.eps} 
\includegraphics[width=66mm]{tXehrehAehVeh5-half2.eps} 
\caption{\small{The comoving distance, proper distance, area and volume of the cosmological event horizon is shown for three different cosmological models.  The models' matter (energy) density and cosmological constant $\omol$ is given in the legend in the upper right corner.  The dimensionless comoving distance is not shown for the $\omol=(0.3,0.7)$ case since $R_0$ is undefined in this model.  Note that although the radius and volume within the cosmological event horizon both decrease for periods in the $\omol=(0.3,1.4)$ universe, the area always increases.}}
\label{fig:tXehrehAehVeh}
\ectr\end{figure}
\begin{figure}\bctr
\includegraphics[width=66mm]{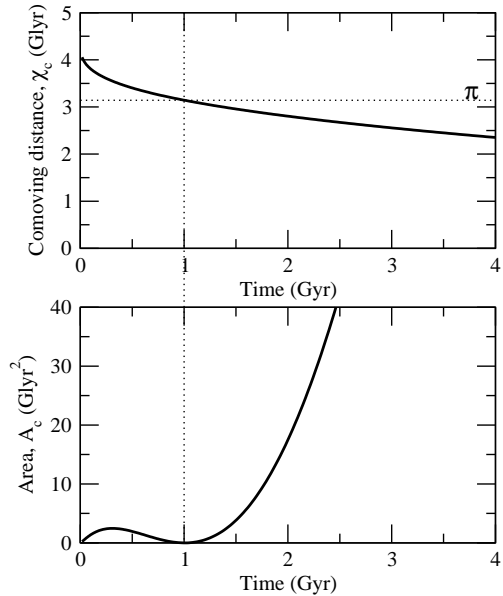}  
\caption{\small{This is a close-up of the region near the origin of Fig.~\ref{fig:tXehrehAehVeh} for the $\omol=(0.3,1.4)$, $k=+1$ case, which appears to show a curious rise and fall in event horizon area at early times.  However, this is an artefact of the finite spatial size of closed FRW universes.  When the comoving distance to the event horizon exceeds $\pi$ it is possible for an observer to see past the antipode.  The event horizon appears at the antipode at the finite time $1.0\, Gyr$ in our example.
}}
\label{fig:tXehrehAehVeh-zoom}
\ectr\end{figure}

The time dependence of the scalefactor, $R(t)$, is given by the Friedmann equations,
\bea \dot{\rho} &=& -3H(\rho + p)\label{eq:Fried1},\\
3H^2&=&8\pi\rho + \Lambda - 3k/R^2\label{eq:Fried2},\eea 
where $\rho$ and $p$ are the density and pressure of the cosmological fluid respectively and $H=\dot{R}/R$ is Hubble's constant.  We assume the present day Hubble's constant $H_0=70\, kms^{-1}Mpc^{-1}$ throughout.  
The radiation density and cosmological constant can be normalized to $\om =  8\pi \rho_0/3H_0^2$ and $\oll = \Lambda/3H_0^2$ respectively so that $\om+\oll = 1$ represents flat space at the present day.  The dimensionless scalefactor $a(t)$ is defined as $a(t) = R(t)/R_0$ where $R_0$ is the present day radius of curvature of the Universe,
\beq R_0 = \frac{c}{H_0}\; \left|\frac{1}{1-\om-\oll}\right|^{1/2}.\eeq  
Equation~\ref{eq:Fried2} can then be rewritten as,
\beq \dot{a}=H_0\left[1+\om(1/a-1) + \oll(a^2-1)\right]^{1/2}.\eeq
Eternally expanding models possess event horizons if light can not travel more than a finite distance in an infinite time,
\beq \chi_c(t) = \int_t^\infty \frac{dt^\prime}{a(t^\prime)} < \infty. \label{eq:chi}\eeq
Our cosmological event horizon is the distance to the most distant event we will ever see (the distance light can travel between now and the end of time) in contrast to our particle horizon, which is the distance to the most distant object we can {\em currently} see (the distance light has travelled since the beginning of time).
The integral in Eq.~\ref{eq:chi} represents the comoving distance to a comoving observer's cosmological event horizon at time $t$.  
The proper distance to the cosmological event horizon is then $r_{\rm c} = R(t)\chi_{\rm c}$.  The area of the cosmological horizon generalized to curved space is, 
\beq A_{\rm c}=4\pi R^2(t)S_k^2(\chi_{\rm c}),\label{eq:Ac}\eeq
which reduces to $A_{\rm c}=4\pi r_{\rm c}^2$ in flat space.  Gibbons and Hawking (1977) suggested that the entropy of the cosmological event horizon is $A_{\rm c}/4$, analogous to the black hole case (Eq.~\ref{eq:Sbh}).

Davies (1988) showed that the cosmological event horizon area of a FRW universe never decreases, assuming the dominant energy condition holds, $\rho+p\ge 0$.  This is analogous to Hawking's area theorem for black holes (Hawking 1972).  In black holes the dominant energy condition is violated by quantum effects, allowing black holes to evaporate and shrink.  There is no analogous shrinking in cosmological horizon area known.

It is interesting to note that the area of the cosmological event horizon increases even in models in which the radius of the event horizon decreases.  Closed eternally expanding universes have a decreasing event horizon radius at late times, but the effect of curvature forces the area to increase nevertheless, e.g. $\omol=(0.3,1.4)$ in Fig.~\ref{fig:tXehrehAehVeh}.

\section{Radiation filled universe} \label{sect:radFRW}
To investigate the interplay of entropy exchange between the cosmological event horizon and an environment we consider an eternally-expanding FRW universe with a positive cosmological constant, filled with radiation of temperature $T(t)$.  Such a universe has an event horizon radius that tends toward the de Sitter value, $r_{\rm deS} = 1/H$, at late times.   Most $\Lambda>0$ universes tend toward de Sitter at late times except the few that have a large enough energy density to begin recollapse before they become cosmological constant dominated.  We include constants in this and subsequent sections to explicitly ensure environment and horizon entropy are being compared in the same units.  The entropy of the cosmological event horizon is, 
\beq S_{\rm c} = \left(\frac{k c^3}{\hbar G}\right) \frac{A_{\rm c}}{4}. \eeq
\begin{figure}\bctr
\includegraphics[width=60mm]{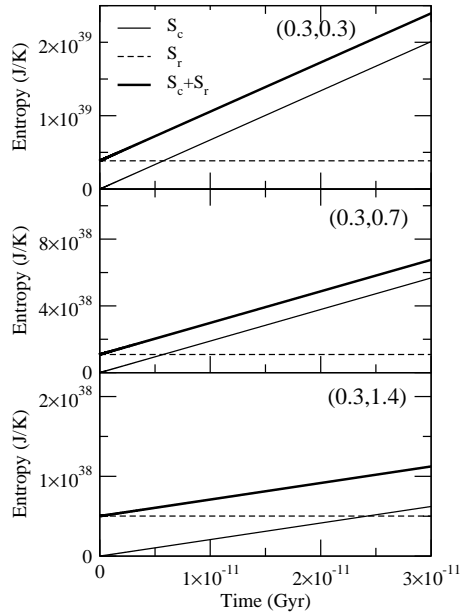}  
\caption{\small{This shows the radiation entropy $S_{\rm r}$ compared to the horizon entropy $S_{\rm c}$ in three radiation filled FRW universes.  Each graph is labeled with the model, $\omol$.  Only early times are shown because that is the only time that the radiation entropy is comparable to the horizon entropy.  The radiation entropy is not constant but decreases rapidly.  However, the decrease is orders of magnitude slower than the increase in cosmological event horizon entropy, so does not show up on this scale.  Total entropy $S_{\rm c} + S_{\rm r}$ never decreases so the GSL holds for these models.}}
\label{fig:Srad}
\ectr\end{figure}
%
Radiation energy density obeys $\rho_{\rm r} = \sigma T^4$ (where the radiation constant $\sigma=\pi^2k^4/15c^3\hbar^3$) 
 while entropy density follows $s_{\rm r}=(4/3)\rho_{\rm r}\, T^{-1}$.  This means the total entropy  within an event horizon volume, $S_{\rm r}=s_{\rm r}V_{\rm c}$, is given by,
\beq S_{\rm r}= \frac{4}{3}\,\sigma^{1/4}\,\rho_{\rm r}^{3/4}\;V_{\rm c}.\eeq
The equations for the volume of the cosmological event horizon in various FRW models are shown in~\ref{app:volume}.   We take $p = \rho_{\rm r}/3$ for radiation in the Friedmann equations (Eq.~\ref{eq:Fried1} and Eq.~\ref{eq:Fried2}).
The radiation density decays as $\rho_{\rm r} = \rho_0 a^{-4}$ (or $T\propto 1/a$) as the universe expands so the radiation entropy within a constant comoving volume ($V\propto a^3$) remains constant.  However, the radiation entropy within the cosmological horizon decreases as the comoving volume of the event horizon decreases ($\chi_c$ decreases in Eq.~\ref{eq:volume}) and radiation crosses the cosmological event horizon.

The evolution of the universe is dependent on the density of radiation, so the model universe we choose constrains the radiation density according to $\Omega_{\rm r} =  8\pi G \rho_0/3H_0^2$.
(The normalized radiation density, $\Omega_{\rm r}$, replaces $\om$ in Friedmann's equation with the difference that $\Omega_{\rm r}$ decays as $a^{-4}$.)  Allowing for this constraint we replace the dust of Sect.~\ref{sect:dustFRW} with radiation and calculate the loss of entropy over the cosmological event horizon as the universe evolves.  Although the radiation represents much more entropy than dust, in a realistic cosmological model this entropy is minuscule compared to that of the cosmological event horizon.  At the present day in a $(\Omega_{\rm r},\oll)=(0.3,0.7)$ radiation dominated FRW universe the radiation entropy would be 14 orders of magnitude smaller than the entropy of the horizon.  At early times the event horizon was tiny and the radiation was very hot -- it is only at early times that we could expect the radiation entropy to be significant enough to compete with the increase in event horizon area.  Figure~\ref{fig:Srad} shows some numerical solutions typical of a wide class of radiation-filled models.  In all cases we find that the total entropy increases with time ($\dot{S}_{\rm r}+\dot{S}_{\rm c}>0$) in conformity with our extended interpretation of the generalized second law of thermodynamics.  Davies and Davis (2002) show analytically that thermal radiation crossing the cosmological event horizon satisfies the GSL in the limit of small departures from de Sitter space as long as the radiation temperature is higher than the cosmological horizon temperature.  A rigorous analytical proof for the general FRW case, however, is lacking.

\section{Black hole-de Sitter spacetimes}\label{sect:bhFRW}
A way to directly compare the entropic worth of cosmological horizons and black hole horizons is to assess the change in entropy as black holes cross the cosmological horizon.  To this end we examine FRW universes containing a dilute pressureless gas of equal mass black holes.  We ignore the Hawking effect which would be negligible for black holes larger than solar mass over the timescales we address\footnote{Black hole evaporation time $\sim (m/m_{\rm solar})^3 \times 10^{66} yr$.}.
As the universe expands the density of the black hole gas decreases ($\rho_{\rm b}\propto a^{-3}$) and black holes disappear over the cosmological event horizon, resulting in a decrease in the black hole contribution to the total entropy within a horizon volume.  The area of the cosmological event horizon increases in turn\footnote{Cause and effect become confused when we try to assess cosmological event horizons in an analogous way to black holes.  The normal language used for cosmological event horizons would be to say that the matter density and cosmological constant of the universe determine the rate of expansion of the universe and thus determine the increase in distance to the event horizon.  Alternatively we can state that the loss of matter (energy) over the cosmological horizon results in the increase in distance to the event horizon.}.   To ascertain whether the GSL is threatened we ask: does the cosmological event horizon area increase enough to compensate for the loss of black hole entropy?   

We find that for realistic cosmological models the increase in cosmological horizon area overwhelms the loss of black hole horizon area, in clear conformity with the extended GSL.  Greater interest, then, attaches to the case where the black holes are relatively large enough to represent a significant fraction of the total horizon area.  In a realistic case this would refer only to very early epochs, on the assumption that primordial black hole formation had taken place.  In what follows we concentrate on the case where the ratio of black hole horizon area to total horizon area is large.  

Davies and Davis (2002) show that black holes crossing the cosmological event horizon maintain the GSL in the limit of small departures from de Sitter space as long as $r_{\rm b}\llsim r_{\rm c}$ (the black holes are smaller than the cosmological event horizon).  Here we summarize numerical investigations that extend this work to general cosmological models.

The area of the cosmological event horizon is easy to calculate in arbitrary (eternally expanding) FRW universes, as shown in Sect.~\ref{sect:dustFRW}.  Not so the event horizon area of black holes because the solutions require us to deal with an overdensity in an homogeneous, time-dependent background.  The Schwarzschild metric applies for a black hole embedded in empty space and the relationship between black hole mass and event horizon radius, $r_{\rm b}$, is $m_{\rm b}=r_{\rm b} c^2/2G$.  The Schwarzschild-de Sitter solution applies for a black hole embedded in a de Sitter universe (a universe with zero mass density and a constant positive cosmological constant, $\Lambda$).  This solution should therefore be a better approximation than pure Schwarzschild at late times in a FRW universe with $\Lambda >0$.  The mass of a black hole in such a space is (Gibbons \& Hawking, 1977),
\beq m_{\rm b}= \frac{r_{\rm b} c^2}{2G}\left(1-\frac{\Lambda r_{\rm b}^2}{3c^2}\right). \label{eq:mbsdeS}\eeq
There are two positive real solutions for $r_{\rm b}$.  The outer is identified with the cosmological event horizon radius, $r_{\rm c}$.   We approximate a black hole embedded in an arbitrary FRW universe using the Schwarzschild-de Sitter solution.  At early times black holes would have a smaller horizon area than this approximation due to the presence of other black holes within the cosmological horizon.

\begin{figure*}\bctr
\includegraphics[width=44mm]{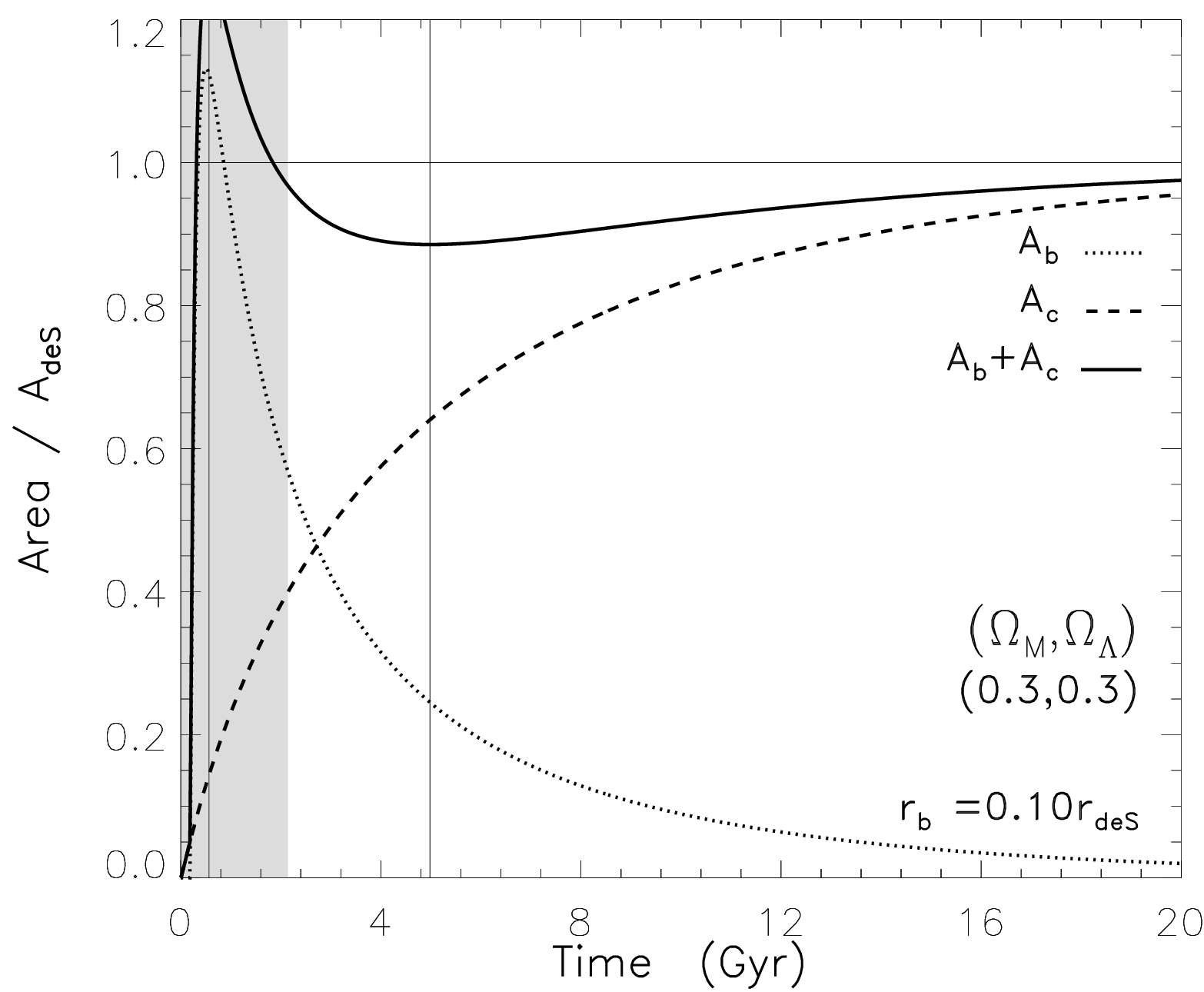}
\includegraphics[width=44mm]{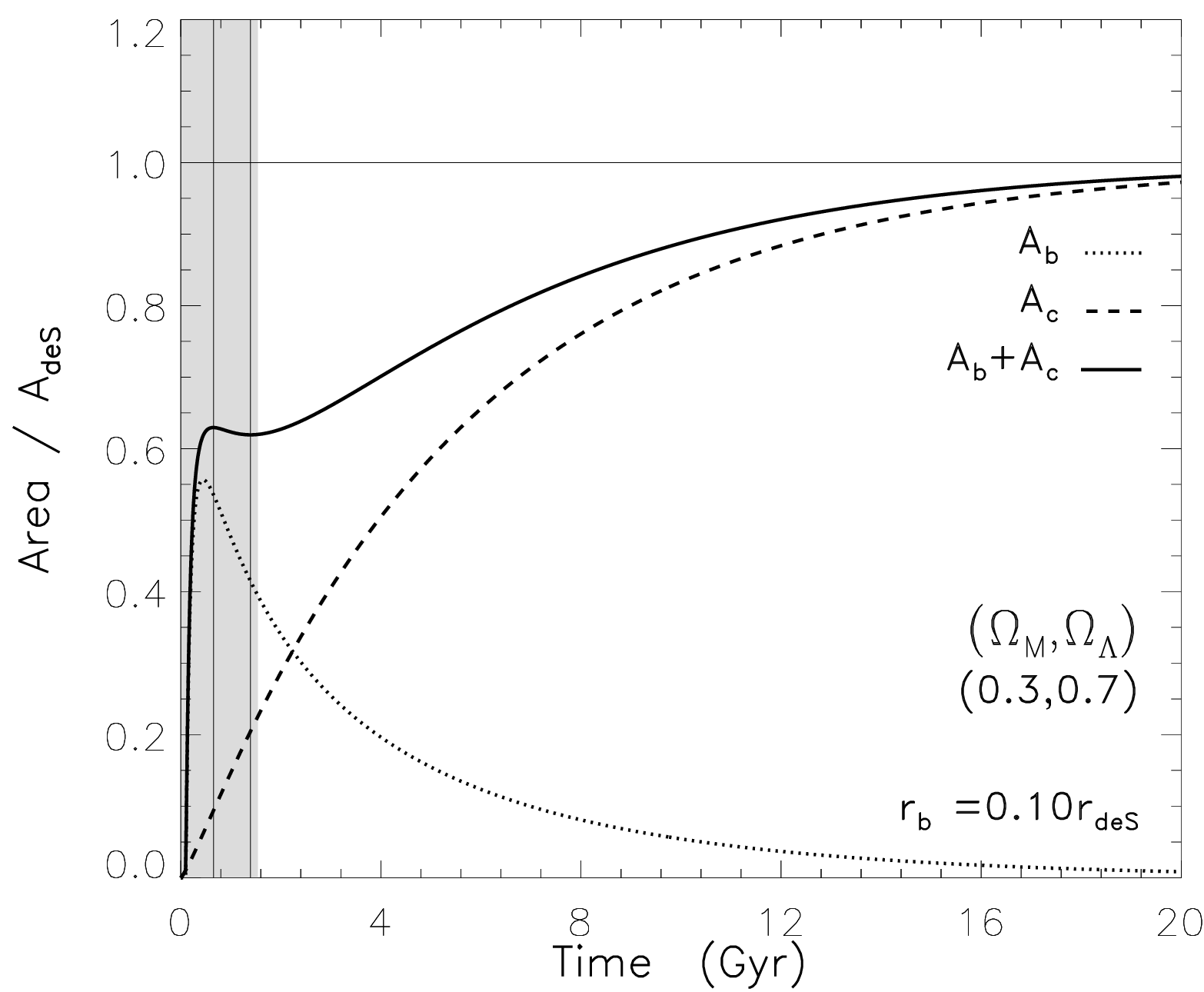}
\includegraphics[width=44mm]{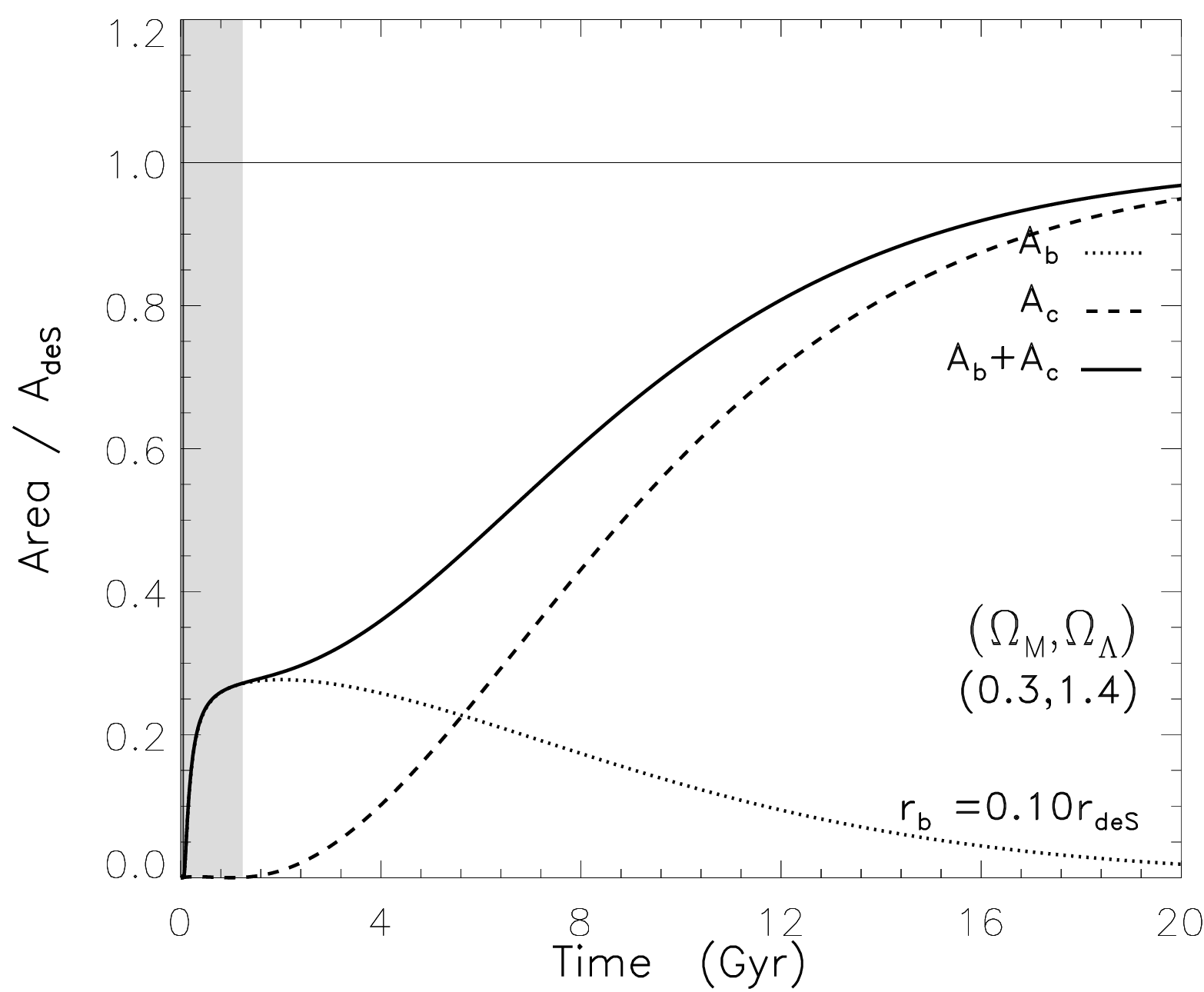}
\includegraphics[width=44mm]{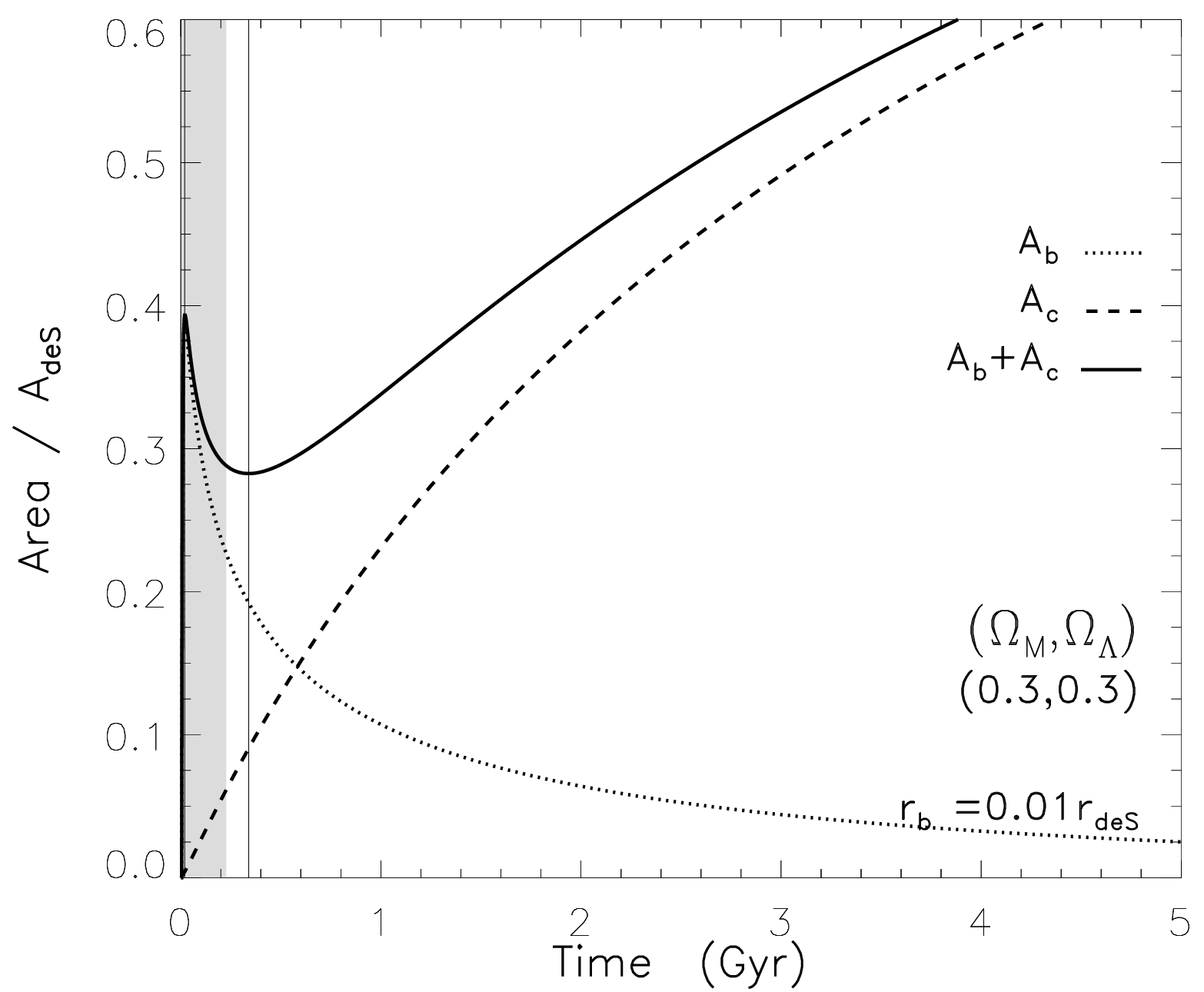}
\includegraphics[width=44mm]{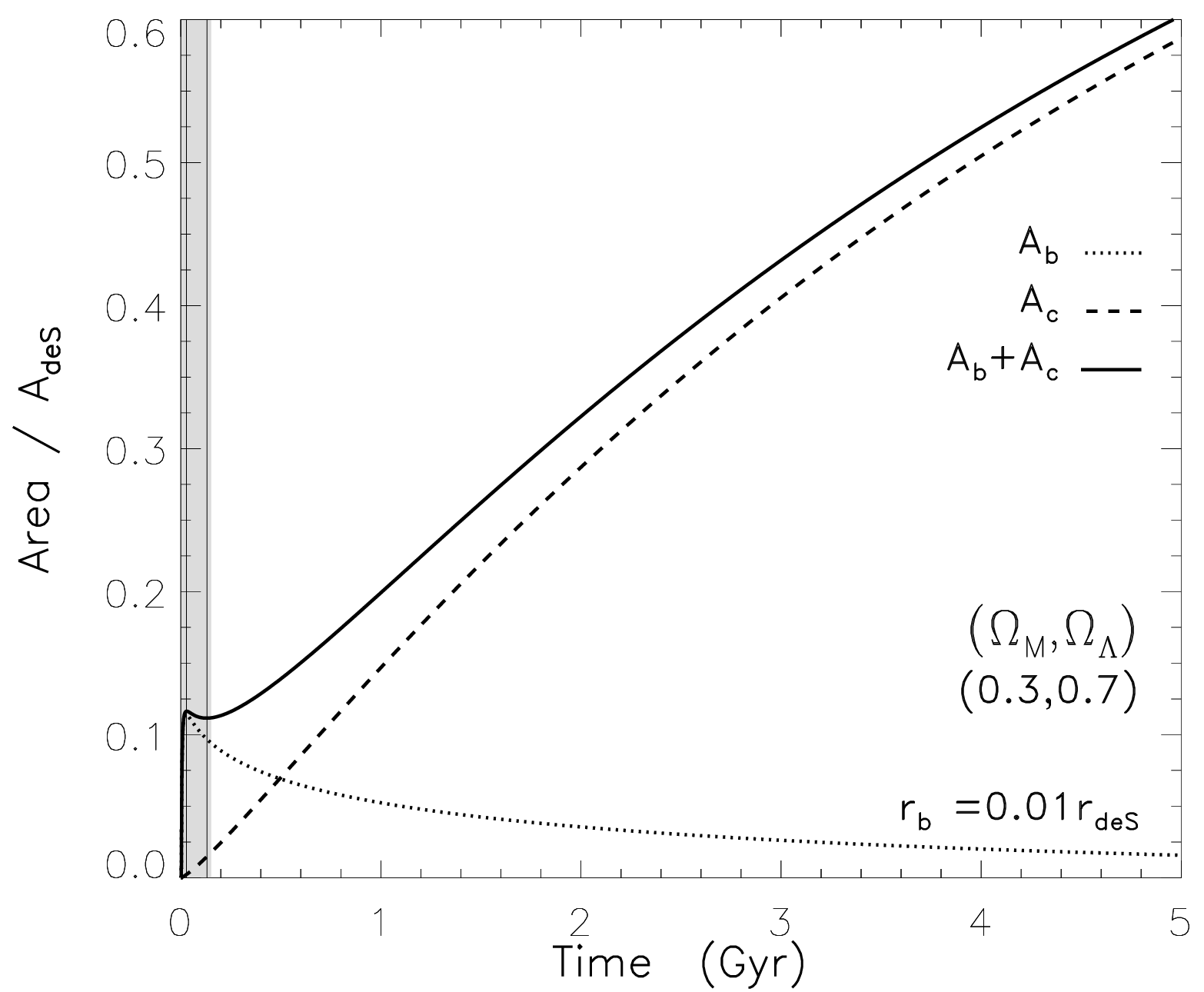}
\includegraphics[width=44mm]{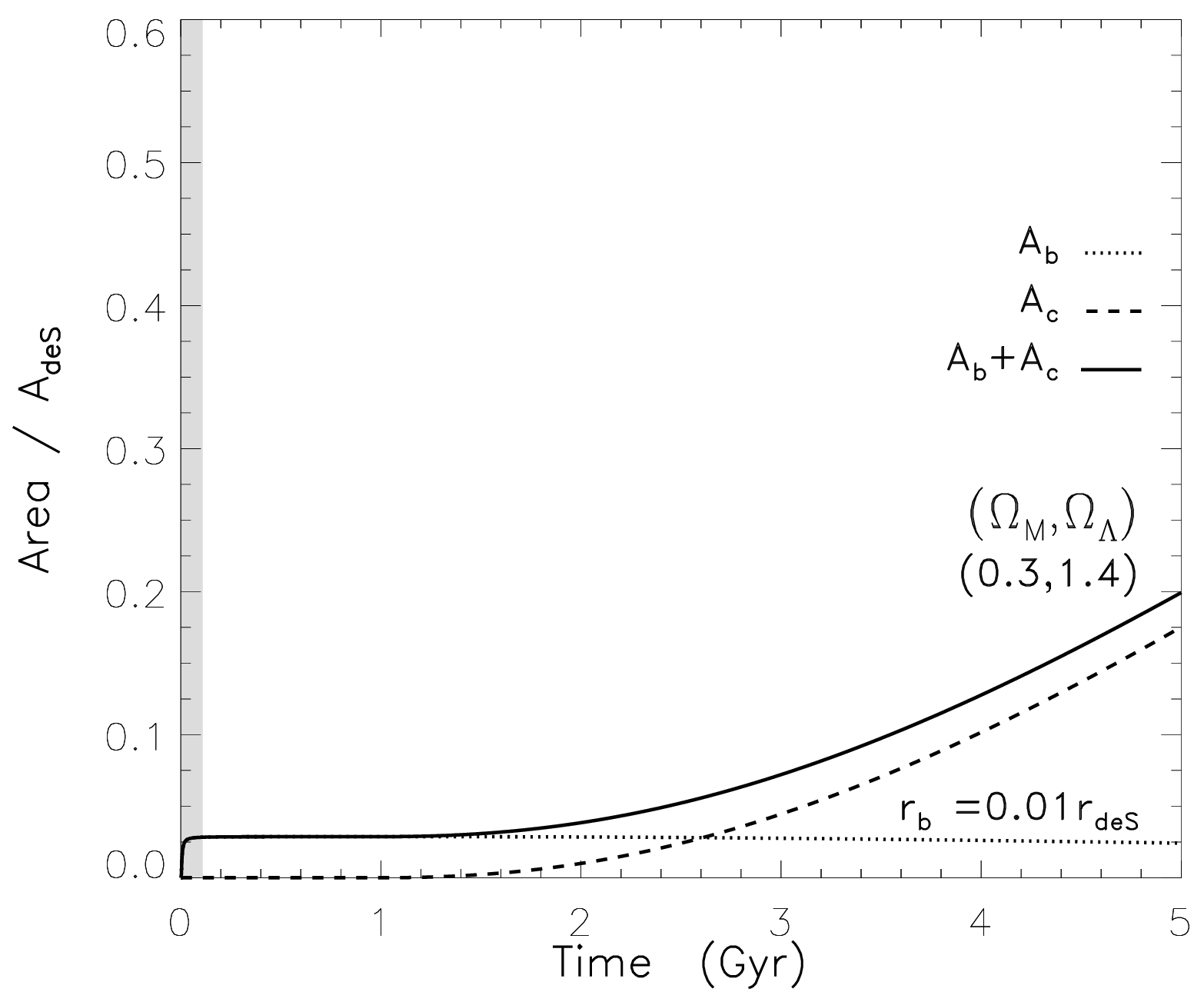}
\caption{The evolution of total horizon area is shown as a function of time for three FRW models filled with a pressureless gas of black holes.  The upper row has $r_{\rm b}=0.1r_{\rm deS}$ while the lower row has $r_{\rm b}=0.01r_{\rm deS}$.   The vertical axis has been scaled to the de Sitter horizon area, $A_{\rm deS}$, in each model.    The dotted line shows the total area of black hole horizons within the cosmological event horizon.  The dashed line shows the area of the cosmological event horizon.  The thick solid line shows the sum of the black hole and cosmological horizon areas.  The thin, solid vertical lines mark turning points in the total horizon area curve.
Corrections have been made for the three assumptions listed in Sect.~\ref{sect:bhFRW}.  The gray shading indicates the region that should be neglected because black holes overlap.  The black hole contribution to area starts from zero and peaks because black holes initially have a radius larger than the cosmological horizon radius and so are excluded from the area calculation by Eq.~\ref{eq:rc}.
Here the areas of black holes have been calculated assuming they were in the geometry of the type of universe they are embedded in (using Eq.~\ref{eq:Abhsingle}). The results are qualitatively unchanged when $A_{\rm b}=4\pi r_{\rm b}^2$ is used.}
\label{fig:area}\ectr
\end{figure*}

We have the freedom to choose the mass of our black holes arbitrarily.  The number density of black holes is then constrained by the need to remain consistent with the matter density of the universe.  
Recall, the normalized matter density of the universe, $\om$, is related to the density by,
\beq      \rho_0=\frac{3H_0^2\,\om}{8\pi G}.\eeq
We assume that the black holes are the only contribution to the matter density of the universe, $\rho_0 = \rho_{\rm b_0}$.  Let $n_{\rm b_0}$ be the current number density of black holes. Then,
\bea \rho_{\rm b_0}&=& m_{\rm b}\;n_{\rm b_0,}\\
n_{\rm b_0}&=&\frac{3H_0^2\,\om}{8\pi\, G\,m_{\rm b}}.\eea
The black hole number density drops like $n_{\rm b}=n_{\rm b_0}a^{-3}$ as the universe expands. 
The surface area of a single black hole's event horizon will depend to some extent on the spacetime geometry of the cosmological model.  For a wide range of values of the ratio $r_{\rm b}/R$ the resulting corrections to the black hole horizon area are negligible.  But for very large black holes or very early epochs these corrections may be significant.  A full treatment of black hole solutions in time-dependent cosmological backgrounds is beyond the scope of this paper.  As a first approximation, however, we may correct for the spacetime curvature of the embedding space by introducing the factor $S_k$ such that,
\beq A_{\rm b}=4\pi R^2(t)\,S_k^2(r_{\rm b}/R),\label{eq:Abhsingle}\eeq
 (c.f. Eq.~\ref{eq:Ac}).  This factor is chosen to make the areas of the black hole and cosmological horizons the same when $r_{\rm b}=r_{\rm c}$. 
Thus the total surface area of all the black hole event horizons, $A_{\rm b,tot}$, is given from Eqs.~\ref{eq:mbsdeS}--\ref{eq:Abhsingle} and Eq.~\ref{eq:volume} by,
\beq 
A_{\rm b,tot}= A_{\rm b}\; n_{\rm b}\; V_{\rm c}\label{eq:Abh}\eeq
We use numerical calculations to find the comoving distance to the cosmological event horizon from which we can calculate both $A_{\rm c}$ (Eq.~\ref{eq:Ac}) and $V_{\rm c}$ (Eq.~\ref{eq:volume}), in turn allowing us to use Eq.~\ref{eq:Abh} for $A_{\rm b,tot}$.  

The de Sitter horizon at $r_{\rm deS}=\sqrt{3/\Lambda}$ is the horizon that would exist if the matter density were zero in each model.  As such it is the asymptotic limit in time of the cosmological event horizon.  We express the results of the numerical calculations in terms of the radius and area of the de Sitter horizon.  The results of these numerical calculations are shown in Fig.~\ref{fig:area}.  Black hole event horizon area, cosmological event horizon area and the total horizon area are plotted against time for a variety of models. 

Treating the problem as stated so far we find significant departures from the GSL at early times in all models and at late times for large black holes.  However, we believe these departures are an artefact of the approximations we have used.
Firstly, by treating the black holes as dilute dust (and as solid spheres) we have neglected interactions between them.  At very early times the black holes in the simulation are so densely packed that they overlap, which is clearly unphysical (see~\ref{app:B}).    Secondly, we have assumed that the disappearance of a black hole across the cosmological horizon is instantaneous, but for black holes of size comparable to the cosmological horizon this is unrealistic.  A proper GR treatment of the merging of horizons, which will involve significant departures from homogeneity and isotropy, is beyond the scope of this paper.  However, as a first approximation to compensating for this effect, we use a simple geometric argument (see~\ref{app:B}).  Taking both the above considerations into account removes almost all the departures from the GSL.
\begin{figure}\bctr
\includegraphics[width=78mm]{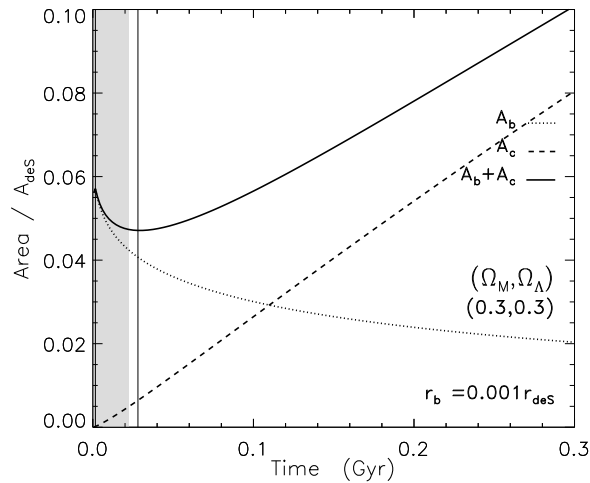}
\caption{An example in which the assumption of the Schwarzschild-de Sitter solution for black hole area breaks down because of the presence of matter density outside the black holes.  The GSL appears to be violated by the entropy decrease at early times even for small black holes.}
\label{fig:area-r0001rdeS-curved}\ectr
\end{figure}

A third approximation which we have used but cannot correct for is the assumption that the Schwarzschild-de Sitter solution for the black hole radius holds.  This neglects the presence of matter density outside the black hole.   This approximation is therefore suspect at early times in FRW universes while the universe is dominated by matter rather than dark energy ($\Lambda$).  An example of a GSL violation which we attribute to the breakdown of the Schwarzschild-de Sitter assumption is shown in Fig.~\ref{fig:area-r0001rdeS-curved} for the spatially open ($k=-1$) model where departures from the GSL are indicated at early times. 

\begin{figure}\bctr
\includegraphics[width=78mm]{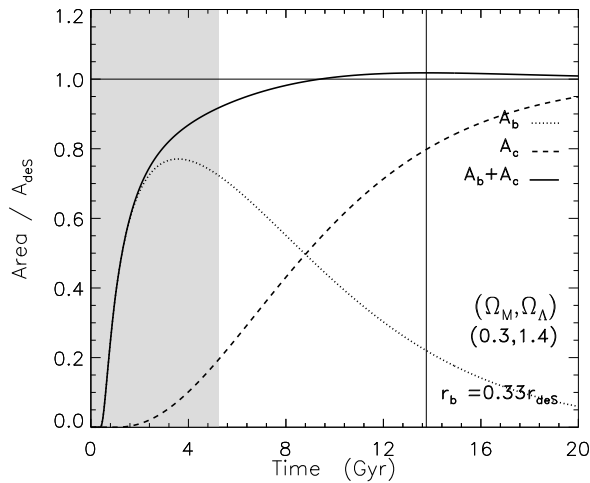}
\caption{An model universe filled with large black holes for which the assumption of the Schwarzschild-de Sitter solution breaks down.  The GSL appears to be violated by the entropy decrease at late times.}
\label{fig:area-r033rdeS-curved}\ectr
\end{figure}

The Schwarzschild-de Sitter approximation also breaks down when the radius of the black hole is comparable to the radius of the cosmological event horizon.
  This is because the effect of the embedding spacetime on the mass-radius relationship of a black hole becomes larger for larger black holes (see the term in brackets in Eq.~\ref{eq:mbsdeS}).    
An example is shown in the spatially closed ($k=+1$) model illustrated in Fig.~\ref{fig:area-r033rdeS-curved}, where departures from GSL are indicated at late times.  

A more accurate resolution of these departures from the GSL awaits the derivation of horizon solutions for black holes embedded in arbitrary FRW spacetimes.
An indication of the magnitude of the effect of different embeddings can be gained by comparing the Schwarzschild-de Sitter solution to the Schwarzschild solution.  For a particular black hole radius the difference in mass for the two embeddings is $\Delta m_{\rm b}/m_{\rm b}=(m^{\rm deS}_{\rm b}-m^{\rm Sch}_{\rm b})/m^{\rm Sch}_{\rm b} = -\Lambda r_{\rm b}^2/3c^2$.  That means that for $H_0=70 kms^{-1}Mpc^{-1}$ and  $\oll = 0.7$ the difference between the two solutions is less than $\Delta m/m = 0.01$ as long as black holes are smaller than 1.7 billion light years across (the de Sitter horizon for a Universe with $\oll=0.7$ sits at $r_{\rm deS}=\sqrt{3/\Lambda}=16.7 Glyr$ so $1.7 Gyr$ represents $r_{\rm b} = 0.10 r_{\rm deS}$, c.f.~Fig.~\ref{fig:area}).   Therefore to minimize the effect of the embedding spacetime on the radius of a black hole we simply need to use ``small'' black holes (a ``small'' black hole of $0.17 Glyr$ radius is still on the order of $10^{21}$ solar masses).

The only GSL violation that does not disappear when black holes are restricted to small sizes is the early time entropy decrease that occurs in open universes because of the breakdown of the Schwarzschild-de Sitter solution in this regime.  
We emphasize that any other apparent departures from GSL are manifested only in the extreme cases where the size of the black holes approach the size of the observable universe.  In a realistic cosmological model, the largest black holes formed by merger will still be orders of magnitude smaller than the cosmological horizon.  In those cosmological models that permit primordial black hole formation from density perturbations, the size of the holes is still generally much less that the cosmological horizon size at the epoch of formation.

\section{Conclusions}
\vspace{-2mm}

We define total entropy to be the entropy of a cosmological event horizon plus the entropy within it.  Davies (1988) showed that the entropy of the cosmological event horizon in FRW universes, subject to the dominant energy condition, never decreases.  We examined radiation filled FRW universes and showed that total entropy never decreases for a wide range of models by testing the parameter space using numerical calculations.  We then assessed the entropy lost as black holes disappeared over the cosmological event horizon.  The lack of a black hole solution for arbitrary spacetime embeddings restricts the application of this technique.  Limiting the size of black holes to those small enough that the difference in embedding in empty space compared to de Sitter space is less than 0.1\% allowed us to show that no GSL violation occurs in any of the closed or flat models tested, but an apparent violation occurs at early times in open FRW universes, probably due to the breakdown of the Schwarzschild-de Sitter assumption in the presence of matter density outside a black hole.  Further progress in resolving this matter will require more realistic approximations of black hole solutions in cosmological backgrounds.  An associated issue that needs to be addressed is what constitutes the appropriate surface that characterises horizon entropy when black holes are situated in a time-dependent background.

\appendix
\section{Volume within cosmological event horizon}\label{app:volume}

The volume within a cosmological event horizon is given by:
\bea V_{\rm c} &=& 4\pi\,R^3  \int_0^{\chi_{\rm c}}S^2_k(\chi)d\chi\\
&=&\left\{ \begin{array}{ll}
	2\pi\,R^3\, (\chi_{\rm c}-\sin\chi_{\rm c}\cos\chi_{\rm c}) & \mbox{closed,} \\ & \\
	\frac{4}{3} \pi R^3 \chi_{\rm c}^3 & \mbox{flat,}\\ & \\
	2\pi\,R^3\, (-\chi_{\rm c}+\sinh\chi_{\rm c}\cosh\chi_{\rm c}) & \mbox{open.}\end{array}\right.\label{eq:volume}
\eea

\section{Geometric considerations}\label{app:B}
We rule out the times when black holes are so close that they overlap as being unphysical.  The separation between black holes is given by $\mbox{separation} = n_{\rm b}^{-1/3}$.  So we rule out any regions for which,
\beq 2r_{\rm b} \le n_{\rm b}^{-1/3}. \label{eq:nb}\eeq
The unphysical region defined by Eq.~\ref{eq:nb} is shaded gray in Figs.~\ref{fig:area}--\ref{fig:area-r033rdeS-curved}.

By considering a black hole to have crossed the cosmological horizon when its centre passes over it we calculate too much black hole horizon area (averaged over all black holes) to be inside the cosmological horizon.
To fix this we need to calculate the point at which exactly half the black hole horizon is outside the cosmological horizon.  This occurs when the black hole's diameter makes a secant to the cosmological horizon.  

\setlength{\unitlength}{5mm}
\begin{figure}[h!]\bctr
\begin{picture}(16,6.5)
	\thicklines
	\put(3,3){\circle{6}} 
	\put(5.11,5.11){\circle{2}} 
	\put(5.11,5.11){\line(-1,1){1.5}}
	\put(5.11,5.11){\line(1,-1){1.5}}
	\put(3,3){\line(1,1){2.1}}
	\put(3.4,3.9){\makebox(0,0){$r_c$}}
	\put(7.1,5.5){\makebox(0,0){black}}
	\put(6.9,4.9){\makebox(0,0){hole}}
	\put(7.5,0.6){\makebox(0,0){cosmological}}
	\put(7.5,0.0){\makebox(0,0){event horizon}}
	\put(5.0,0.0){\vector(-1,1){0.5}}
	\put(10.0,0.0){\vector(1,1){0.5}}
	\put(12,3){\circle{6}} 
	\put(14.4,3){\circle{3.5}} 
	\put(14.4,1.25){\line(0,1){3.5}}
	\put(12,3){\line(1,0){3}}
	\put(12,3){\line(4,3){2.4}}
	\put(13.2,2.3){\makebox(0,0){$r_c-\delta$}}
	\put(12.4,3.8){\makebox(0,0){$r_c$}}
	\put(14.65,2.6){\makebox(0,0){$\delta$}}
	\put(14.0,3.6){\makebox(0,0){$r_{\rm b}$}}
	\put(12,2.8){\vector(1,0){2.4}}
	\put(14.4,2.8){\vector(-1,0){2.4}}
	\put(16.3,4.8){\makebox(0,0){black}}
	\put(16.6,4.2){\makebox(0,0){hole}}
\end{picture}\ectr
\end{figure}

Therefore we should consider black holes to have left the horizon when they are a distance $\delta$ from the horizon where $\delta$ is the length of the perpendicular bisector of the secant between the secant and the perimeter of the event horizon.  That is, when we calculate the volume within which are black holes we should use the radius $r_c-\delta$, 
\beq    r_c - \delta = \sqrt{r_c^2 -r_{\rm b}^2}. \label{eq:rc}\eeq
This corrected calculation is shown in Fig.~\ref{fig:area}.

\ack{TMD acknowledges an Australian Postgraduate Award.  CHL acknowledges fellowship support from the Australian Research Council.  This work was supported in part by a UNSW Faculty Research Grant.}
\section*{Postscript}An inadvertant omission meant no reference to Padmanabhan (2002) appeared in the published work.  We have added the reference in this version.
\section*{References}
\begin{harvard}
\item Birrell N D and Davies P C W 1981 {\it Quantum fields in curved space} (Cambridge: Cambridge University Press)
\item Bekenstein J D 1973 {\it Phys. Rev. D} {\bf 7} 2333--46
\item Bousso R 2002 {\it Rev.Mod.Phys.} {\bf 74} 825--74 
\item Davies P C W 1984 {\it Phys. Rev. D} {\bf 30} 737--42
\item \dash 1988 {\it Class. Quantum Grav.} {\bf 5} 1349--55
\item Davies P C W and Davis T M 2002 {\it Foundations of Phys.} {\bf 32} 1877--1889
\item Davies P C W, Ford L H and Page D N 1984 {\it Phys. Rev. D} {\bf 34} 1700--07
\item Gibbons G W and Hawking S W 1977 {\it Phys. Rev. D} {\bf 15} 2738--51
\item Hawking S W 1972 {\it Comm. Math. Phys.} {\bf 25} 152--166
\item \dash 1975 {\it Comm. Math. Phys.} {\bf 43} 199--220
\item Lloyd S 2002 {\it Phys. Rev. Lett.} {\bf 88} 237901
\item Padmanabhan T 2002 {\it Class. Quantum Grav.} {\bf 19} 5387--5408
\item Susskind L 1995 {\it J. Math. Phys.} {\bf 36} 6377--96
\end{harvard}
\end{document}